  \documentclass{gretsi}
 \usepackage[english,french]{babel}   
  \usepackage{times}			
 
\usepackage[T1]{fontenc}
\usepackage{amsmath, amssymb}
\newcommand{\norm}[1]{\left\lVert#1\right\rVert}
\DeclareMathOperator*{\argmin}{arg\,min}
\DeclareMathOperator*{\argmax}{arg\,max}
\newtheorem{theorem}{Theorem}
\usepackage[ruled, noline, linesnumberedhidden, shortend,french, frenchkw]{algorithm2e}
\usepackage{cite}
\usepackage{graphicx}
\usepackage{subfig}
\usepackage{bm}
\usepackage{url}
\usepackage{microtype}

\usepackage{comment}

\titre{Une version polyatomique de l'algorithme Frank-Wolfe\\pour résoudre le problème LASSO en grandes dimensions}

\auteur{\coord{Adrian}{Jarret}{1},
        \coord{Matthieu}{Simeoni}{2},
        \coord{Julien}{Fageot}{1}}

\adresse{\affil{1}{Laboratoire des Communications AudioVisuelles \\
         Bâtiment BC, Route Cantonale, 1015 Lausanne, Suisse}
         \affil{2}{EPFL Center for Imaging \\
         Bâtiment BM, Station 17, 1015 Lausanne, Suisse}}

\email{adrian.jarret@epfl.ch, matthieu.simeoni@epfl.ch, julien.fageot@epfl.ch}

\resumefrancais{
Nous nous intéressons à la reconstruction parcimonieuse d'images à l'aide du problème d'optimisation régularisé LASSO. Dans de nombreuses applications pratiques, les grandes dimensions des objets à reconstruire limitent, voire empêchent, l'utilisation des méthodes de résolution proximales classiques. C'est le cas par exemple en radioastronomie. Nous détaillons dans cet article le fonctionnement de l'algorithme \textit{Frank-Wolfe Polyatomique}, spécialement développé pour résoudre le problème LASSO dans ces contextes exigeants. Nous démontrons sa supériorité par rapport aux méthodes proximales dans des situations en grande dimension avec des mesures de Fourier, lors de la résolution de problèmes simulés inspirés de la radio-interférométrie.
}

\resumeanglais{We consider the problem of recovering sparse images by means of the penalised optimisation problem LASSO. For various practical applications, it is impossible to rely on the proximal solvers commonly used for that purpose due to the size of the objects to recover, as it is the case for radio astronomy. In this article we explain the mechanisms of the \textit{Polyatomic Frank-Wolfe algorithm}, specifically designed to minimise the LASSO problem in such challenging contexts. We demonstrate in simulated problems inspired from radio-interferometry the preeminence of this algorithm over the proximal methods for high dimensional images with Fourier measurements.}

\begin{document}
\maketitle

\let\thefootnote\relax\footnotetext{Ce travail a été financé par le Fonds national Suisse (FNS) à travers les bourses CRSII5\_193826  AstroSignals (A. Jarret and M. Simeoni), 200 021 181 978/1, “SESAM - Sensing and Sampling: Theory and Algorithms” (M. Simeoni) et P400P2\_194364 (J. Fageot).}

\section{Introduction}

La reconstruction parcimonieuse de signaux ou d'images présente de nombreux intérêts, que ce soit sur le plan pratique ou théorique. Entre autres, elle induit souvent une sélection de variables, ce qui permet d'obtenir des modèles interprétables avec un plus grand pouvoir de généralisation. Utiliser une hypothèse de parcimonie \textit{a priori} peut également aider à modéliser certains phénomènes physiques. L'application qui nous intéresse dans cet article est la reconstruction de sources lumineuses ponctuelles sur des images du ciel en radio-interférométrie pour l'astronomie. Une méthode encore largement utilisée pour ce problème est donnée par l'algorithme CLEAN \cite{Hogbom_1974}, qui met l'accent sur la parcimonie en produisant une solution par l'ajout successif de composants, appelés \textit{atomes}, issus d'un dictionnaire de reconstruction.


Une autre méthode, désormais classique pour la reconstruction parcimonieuse en traitement du signal, consiste à résoudre un problème inverse sous la forme d'un problème d'optimisation pénalisé, dont le terme de pénalité induit de la parcimonie dans les solutions. Nous nous focalisons sur le problème suivant, connu sous le nom de LASSO ou \textit{Basis Pursuit}:
\begin{equation}
    \label{eq:lasso}
    \argmin_{\mathbf{x}\in\mathbb{R}^N} {\frac{1}{2}\norm{\mathbf{y} - \mathbf{Gx}}_2^2 + \lambda \norm{\mathbf{x}}_1 },
\end{equation}
dans lequel $\mathbf{G}$ représente la modélisation linéaire d'un système d'acquisition et $\mathbf{y}\in\mathbb{C}^L$ est supposé être le résultat bruité de la mesure par $\mathbf{G}$ d'un certain signal source $\mathbf{x}_0$, de telle sorte que $\mathbf{y} \approx \mathbf{Gx}_0$. Le paramètre $\lambda$ présente l'avantage de pouvoir contrôler l'équilibre entre fidélité avec les données et parcimonie de la solution. En effet, avec ce choix de pénalité, il a été prouvé l'existence de solutions au plus $K$-parcimonieuses, avec $K$ inférieur au nombre de mesures $L$ \cite[Théorème 6]{unser_representer_2016}.

Cette dernière approche n'a cependant reçu que peu de considération en radio-interférométrie. En effet, les principales méthodes actuelles de résolutions du LASSO, telles que APGD \cite{parikh2014proximal} ou FISTA \cite{beck_fast_2009}, sont basées sur des algorithmes proximaux et manipulent des solutions intermédiaires denses. Or, la radioastronomie requiert d'utiliser des données de très grande dimension, rendant ainsi prohibitif le coût en mémoire de ces méthodes.



Dans cet article, nous présentons le fonctionnement de l'algorithme \textbf{Frank-Wolfe Polyatomique}, récemment proposé dans \cite{Jarret_Fageot_Simeoni_2022}, pour résoudre le problème LASSO. En conservant une structure parcimonieuse pour ses solutions intermédiaires, cet algorithme est particulièrement adapté lorsque la taille des problèmes limite l'usage des méthodes usuelles. Nous étendons les cadres d'application dans lesquels cet algorithme démontre sa supériorité, en le confrontant à des problèmes d'optimisation simulés mettant en jeu des mesures de Fourier. Ces situations, inspirées de la radioastronomie, sont numériquement exigeantes à la fois par leur dimension et la nature complexe des données manipulées.

\section{Frank-Wolfe Polyatomique}

Afin de présenter le fonctionnement de l'algorithme \textit{Frank-Wolfe Polyatomique} (P-FW), nous détaillons d'abord sur la version classique, souvent appelée \textit{Vanilla Frank-Wolfe (V-FW)}.


\paragraph{Frank-Wolfe, un algorithme glouton.}

L'algorithme de Frank-Wolfe (FW) \cite{frank_algorithm_1956}, dans sa version originale, résout un problème d'optimisation de la forme
\begin{equation}
    \label{eq:obj_fw}
    \argmin_{\mathbf{x} \in \mathcal{D}}f(\mathbf{x})
\end{equation}
avec $f$ une fonction de coût convexe et continûment différentiable et $\mathcal{D}$ un sous-ensemble compact convexe d'un espace de Banach.

Il est connu que le problème LASSO \eqref{eq:lasso} peut être reformulé sous la forme \eqref{eq:obj_fw}, en le modifiant légèrement pour le rendre différentiable et en introduisant l'espace de recherche
$$\mathcal{D} = \{\mathbf{x}\in\mathbf{R}^N : \norm{\mathbf{x}}_1 \leq M \}$$
avec $M = \norm{\mathbf{y}}_2^2/2\lambda$ \cite{denoyelle_sliding_2020, harchaoui2012conditional}. On obtient ainsi l'algorithme d'optimisation V-FW fourni en Algorithme~\ref{alg:vfw}. La quantité $\bm{\eta}_k = (1/\lambda) \mathbf{G}^*(\mathbf{y} - \mathbf{Gx}_k)$ est appelée \textit{certificat dual empirique} et $\mathbf{e}_i$ représente le $i$-ème vecteur de la base canonique de $\mathbb{R}^N$.
\begin{algorithm}[h!]
\DontPrintSemicolon
\caption{V-FW pour le LASSO}
\label{alg:vfw}
Initialisation:  $\mathbf{x}_0 = \mathbf{0}$ \;
\Pour{$k=1, 2 \cdots$}{
    \nlset{1)\hspace{-3pt}} \label{algstep:gradient}Déterminer une direction d'avancement:\\
    \ \ $\mathbf{s}_k = \pm M \mathbf{e}_{i_k} \in \argmax_{\mathbf{s} \in \mathcal{D}}{ \langle \bm{\eta}_k, \mathbf{s} \rangle }$ \;

    \nlset{2.a)\hspace{-3pt}} \label{algstep:step_size}Taille du pas: $\gamma_k \gets \frac{2}{k+2}$\;
    \nlset{2.b)\hspace{-3pt}} \label{algstep:reweighting}Pondération: $\mathbf{x}_{k+1} \gets (1-\gamma_k) \mathbf{x}_k + \gamma_k \mathbf{s}_k $\;
}
\end{algorithm}

Plus en détail, cet algorithme se compose de deux étapes, identifiées dans Algorithme~\ref{alg:vfw} par 1), la recherche de nouvelle direction de descente, et 2.a)-b), l'estimation de la nouvelle solution intermédiaire, comme combinaison convexe de la solution actuelle et de cette direction de descente. On désigne ainsi les directions de descente $\mathbf{s}_k$ identifiées à l'étape 1) en tant qu'\textit{atomes}. Le comportement \textit{glouton} des algorithmes FW provient de la stratégie de sélection de ces atomes, en minimisant l'approximation au premier ordre de la fonction objectif évaluée en la position actuelle.

L'algorithme V-FW converge en terme de fonction objectif à la vitesse $\mathcal{O}(1/k)$, où $k$ est le nombre d'itérations \cite{jaggi_revisiting_2013}, ce qui est inférieur aux méthodes proximales actuelles. En contraste, APGD converge à la vitesse $\mathcal{O}(1/k^2)$ \cite{parikh2014proximal}. Les méthodes FW présentent cependant la propriété de conserver au cours des itérations des solutions intermédiaires parcimonieuses, comme combinaison linéaires d'atomes, et de fait peuvent se révéler performantes en grandes dimensions, quand les algorithmes proximaux sont ralentis par leurs solutions intermédiaires denses (besoins en mémoire plus importants, calculs plus longs).

\paragraph{Résoudre le problème LASSO avec un variant polyatomique.}

Le variant \textit{Frank-Wolfe Polyatomique} présenté dans \cite{Jarret_Fageot_Simeoni_2022} a été conçu pour tirer profit au maximum du comportement glouton de V-FW, et ainsi identifier rapidement les atomes importants dans la construction d'une solution au LASSO, tout en conservant des solutions intermédiaires parcimonieuses, pour rester applicable en grandes dimensions. La procédure numérique est présentée en Algorithme~\ref{alg:pfw}\textsuperscript{1}\footnotetext{\textsuperscript{1}Une version plus détaillée et reproductible de l'algorithme est fournie dans \cite{Jarret_Fageot_Simeoni_2022}.}, avec $\delta>0$ un paramètre de reconstruction à adapter aux données considérées.


\begin{algorithm}[t!]
\DontPrintSemicolon
\caption{FW Polyatomique pour le LASSO}
\label{alg:pfw}
Initialisation: $\mathbf{x}_0 \gets 0, \mathcal{S}_0 \gets \emptyset$ \;
\Pour{$k=1, 2 \cdots$}{
    $\gamma_k \gets 2/(k+2)$\;
    \nlset{1'.a)\hspace{-5pt}}\label{algstep:multi}Exploration polyatomique:\;
    \qquad $\mathcal{I}_k = \left\{ 1 \leq j \leq N : |\bm{\eta}_k[j]| \geq \norm{\bm{\eta}_k}_\infty - \delta\gamma_k\right\}$\;
    \nlset{1'.b)\hspace{-5pt}} Actualisation des indices actifs : $\mathcal{S}_{k} \gets \mathcal{S}_{k-1} \cup \mathcal{I}_k$\;
    \nlset{2'.a)\hspace{-5pt}} Réglage du seuil de précision : $\varepsilon_{k} = \varepsilon_0\gamma_k$\;
    \nlset{2'.b)\hspace{-5pt}}\label{algstep:ista}Évaluation des poids actifs :\;
    \quad $\mathbf{x}_{k+1} \gets \argmin {\frac{1}{2}\norm{\mathbf{y} - \mathbf{G} \mathbf{x}}_2^2 + \lambda \norm{\mathbf{x}}_1 }$ \\
    \vspace{3pt} avec la contrainte $\mathrm{Supp}(\mathbf{x})\in \mathcal{S}_k$,\\ 
    résolution interrompue au seuil de précision $\varepsilon_k$
}
\end{algorithm}

Le coeur de l'algorithme P-FW réside dans l'étape 1'.a), au cours de laquelle, contrairement à FW original, il est possible de rajouter plusieurs atomes à la solution intermédiaire actuelle. Les atomes identifiés par P-FW ayant la même forme que ceux identifiés par V-FW, considérer plusieurs atomes candidats revient à autoriser la solution intermédiaire à accéder à de nouvelles coordonnées d'indices $\mathcal{I}_k$, non visitées jusque là. L'ensemble des coordonnées accessibles après cette mise-à-jour à l'étape $k$ est noté $\mathcal{S}_k$ (étape 1'.b)\ ).

Au cours de l'étape 2'.b), l'algorithme attribue un poids aux derniers atomes ajoutés et ajuste les poids des atomes précédents. Pour cela, il ré-optimise la fonction objectif du LASSO en ne considérant que les solutions dont le support est contraint à vivre parmi les coordonnées identifiées actives $\mathcal{S}_k$. Cela revient à résoudre un nouveau sous-problème LASSO, dont la dimension est extrêmement réduite par rapport au problème initial. Numériquement, nous utilisons l'algorithme d'optimisation ISTA \cite{beck_fast_2009}, initialisé avec la solution intermédiaire actuelle. De plus, pour réduire le temps de calcul et tenir compte du fait que la solution partielle évolue beaucoup d'une itération à l'autre aux cours des premières itérations, nous arrêtons cette procédure d'estimation des poids prématurément, à l'aide du critère d'arrêt $\varepsilon_k$. Ce critère d'arrêt est décroissant pour permettre, à terme, une convergence fine de P-FW.

\paragraph{Convergence.}

En s'appuyant sur les résultats présentés dans l'article \cite{jaggi_revisiting_2013}, nous sommes capables de garantir que la séquence des solutions intermédiaires produites par P-FW converge en terme de fonction objectif vers un minimum du problème LASSO. La vitesse de convergence théorique est du même ordre que pour l'algorithme FW classique \cite[Théorème~1]{Jarret_Fageot_Simeoni_2022}.
\begin{theorem}[Convergence de P-FW]
Soit $\mathbf{x}_k$ la suite des solutions intermédiaires produites par P-FW. Notons $\mathcal{L}(\mathbf{x}) = \frac{1}{2}\norm{\mathbf{y} - \mathbf{Gx}}_2^2 + \lambda \norm{\mathbf{x}}_1$ la fonction objectif du problème LASSO et $\mathcal{L}^*$ sa valeur optimale.  Nous avons alors que
\begin{equation*}
    \mathcal{L}(\mathbf{x}_k) - \mathcal{L}^* = \mathcal{O}\left(\frac{1}{k} \right)
\end{equation*}
\end{theorem}


Dans nos expériences (Section~\ref{sec:resultats}), nous observons que la convergence est souvent plus efficace que APGD dans les régimes considérés.

\section{Implémentation}

Afin d'évaluer les performances de P-FW, nous le comparons à APGD pour résoudre le problème LASSO sur des données simulées. Les simulations sont réalisées en Python, à l'aide de la librairie d'optimisation \texttt{Pycsou} \cite{SIMEONI_Pla_2021}. Cette dernière permet de résoudre facilement des problèmes inverses à l'aide d'algorithmes proximaux puissants. Le logiciel implémente de façon très modulaire les principaux éléments constituants de problèmes génériques d'optimisation convexe (fonctionnelles de coût, termes de pénalisation et opérateurs linéaires). En particulier, pour utiliser les algorithmes fournis par \texttt{Pycsou}, il est nécessaire de définir l'application directe et l'application de l'adjoint de l'opérateur de mesure, liant les données à l'inconnue du problème inverse.

Nous détaillons dans cette section les subtilités à prendre en compte pour l'implémentation. 

\paragraph{Opérateur de mesure.}

Pour nos simulations, nous utilisons comme opérateur de mesure un sous-échantillonnage aléatoire de la transformée de Fourier discrète (2D) de l'image d'entrée. Ce choix nous place dans un cadre d'étude similaire à celui de la radio-interférométrie, dont l'opérateur est généralement modélisé par des mesures de Fourier (non uniformes en fréquence).

Soit une image $\mathbf{x} \in \mathbb{R}^N$ de taille $N = n \times n$ et $L$ coordonnées fréquentielles $(u_\ell, v_\ell) \in [0, n-1]^2$,  l'opérateur de mesure est donné par
\begin{equation}
\begin{split}
        \left(\mathbf{Gx}\right)[\ell] &= \frac{1}{n} \sum_{p=0}^{n-1} \sum_{q=0}^{n-1} \mathbf{x}[p, q] \exp{\left(-j \frac{2\pi}{n}(u_\ell p + v_\ell q)\right)} \\ 
        &= \mathrm{DFT}_\mathrm{2D}(\mathbf{x})[u_\ell, v_\ell].
\end{split}
\end{equation}

Par définition, cet opérateur est à valeurs complexes, sa signature est donnée par $\mathbf{G} : \mathbb{R}^N \to \mathbb{C}^L$. L'optimisation convexe faisant l'hypothèse d'espaces vectoriels réels, il convient de définir l'adjoint $\mathbf{G}^* : \mathbb{C}^L \to \mathbb{R}^N$ par rapport au produit scalaire hermitien sur  $\mathbb{C}^L$ :
\begin{equation*}
\begin{split}
    \forall \mathbf{a}, \mathbf{b} \in \mathbb{C}^L, \quad \langle \mathbf{a}, \mathbf{b} \rangle_{\mathbb{C}^L} &= \Re\left( \overline{\mathbf{a}}^T \mathbf{b} \right) 
\end{split}
\end{equation*}
où $\Re : \mathbb{C} \to \mathbb{R}$ désigne la partie réelle d'un nombre complexe. On obtient ainsi l'expression de l'opérateur adjoint à implémenter :
\begin{equation}
    \forall \mathbf{z} \in \mathbb{C}^L, \quad \mathbf{G}^* \mathbf{z} =  \Re\left( \overline{\mathbf{G}}^T \mathbf{z} \right)
\end{equation}
    Pour une meilleure lisibilité, les opérations ont ici été définies sur des images en deux dimensions, mais sont en pratique implémentées avec les versions aplaties et unidimensionnelles des mêmes images.

\paragraph{Correction des poids : résoudre un problème LASSO à support réduit avec Pycsou.}
Lors de l'étape 2'.b) de P-FW, les poids sont estimés en cherchant la solution optimale au problème LASSO en limitant l'espace de recherche à l'enveloppe convexe des atomes estimés jusqu'à l'itération actuelle (se référer à \cite{Jarret_Fageot_Simeoni_2022} pour plus de détails). On peut ainsi se ramener à résoudre le problème LASSO à support réduit suivant :
\begin{equation}
    \label{eq:lasso_restreint}
    \argmin_{\widetilde{\mathbf{x}}\in\mathbb{R}^{\mathrm{Card}(\mathcal{S}_k)}} {\frac{1}{2}\norm{\mathbf{y} - \mathbf{G}_{\mathcal{S}_k} \widetilde{\mathbf{x}}}_2^2 + \lambda \norm{\widetilde{\mathbf{x}}}_1 }
\end{equation}

Numériquement, il suffit de récupérer les colonnes d'intérêt de la matrice $\mathbf{G}$, ainsi notée $\mathbf{G}_{\mathcal{S}_k}$, pour contraindre le support de la solution de ce sous-problème. On obtient un problème dont la taille est nettement inférieure au problème initial, peu gourmand en mémoire (en pratique $\mathrm{Card}(\mathcal{S}_k)$ est au maximum de l'ordre de la parcimonie des solutions, d'où $\mathrm{Card}(\mathcal{S}_k) \ll N$). Nous utilisons ensuite l'implémentation de ISTA fournie par \texttt{Pycsou} (à travers la classe \texttt{APGD} et le paramètre \texttt{acceleration=None}) pour obtenir les poids de la prochaine solution intermédiaire.

\section{Résultats}
\label{sec:resultats}

Nous regardons l'évolution de la valeur de la fonction objectif au fil du temps lors de la résolution du problème LASSO. À titre de comparaison, nous reportons les performances de V-FW et de APGD en parallèle de celles obtenues avec P-FW. Quelques corrections mineures ont été intégrées par rapport \cite{Jarret_Fageot_Simeoni_2022}, le code pour reproduire les expériences est accessible en \cite{adriaj}.

\begin{figure}[t!]
\begin{center}
\vspace*{-.5cm}
\subfloat[$K=50, N=101^2$]{
  \includegraphics[trim={0 15 0 0},width=.48\linewidth, clip]{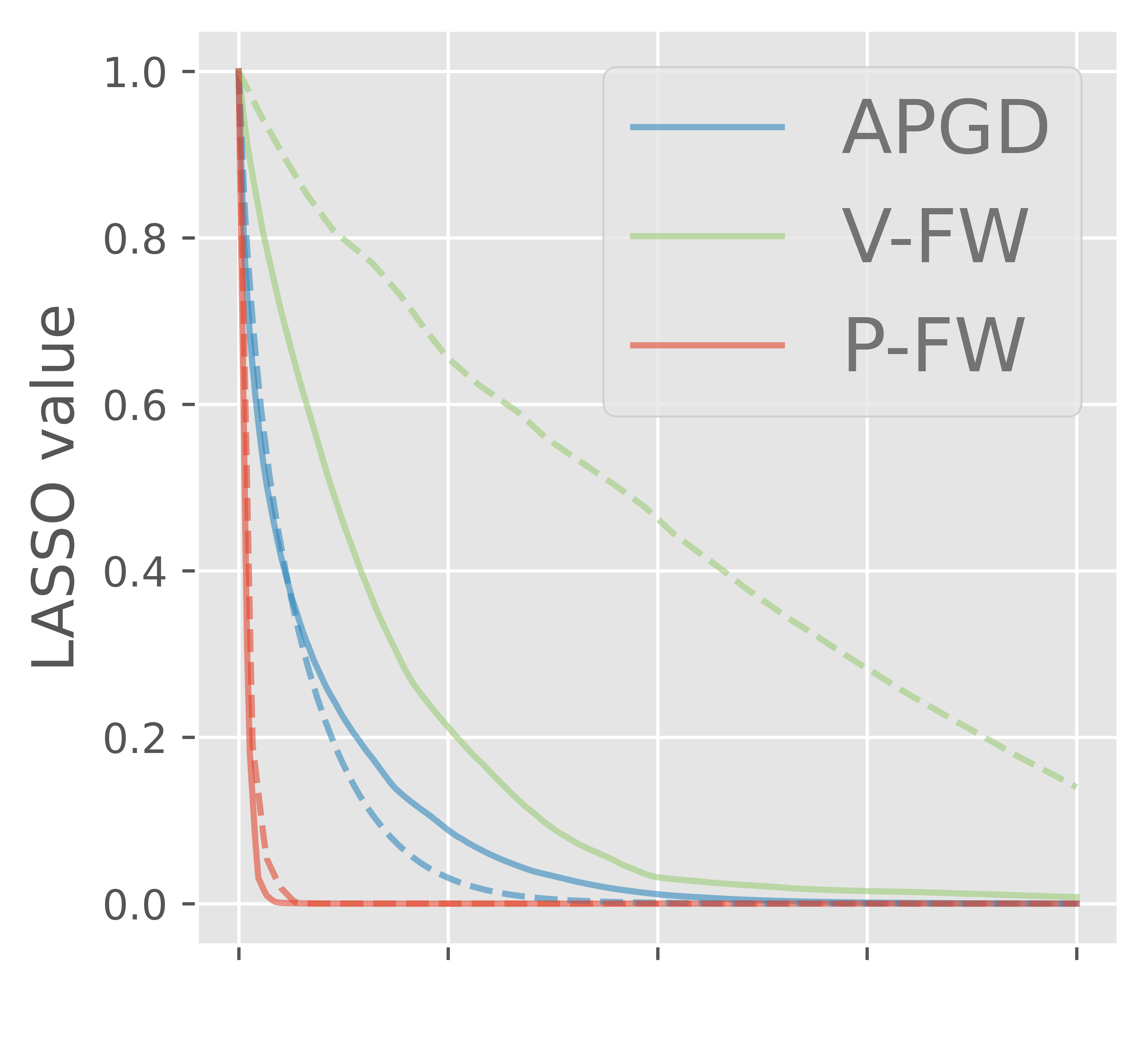}
  }\hfill
\subfloat[$K=50, N=201^2$]{
  \includegraphics[trim={0 15 0 0},width=.45\linewidth, clip]{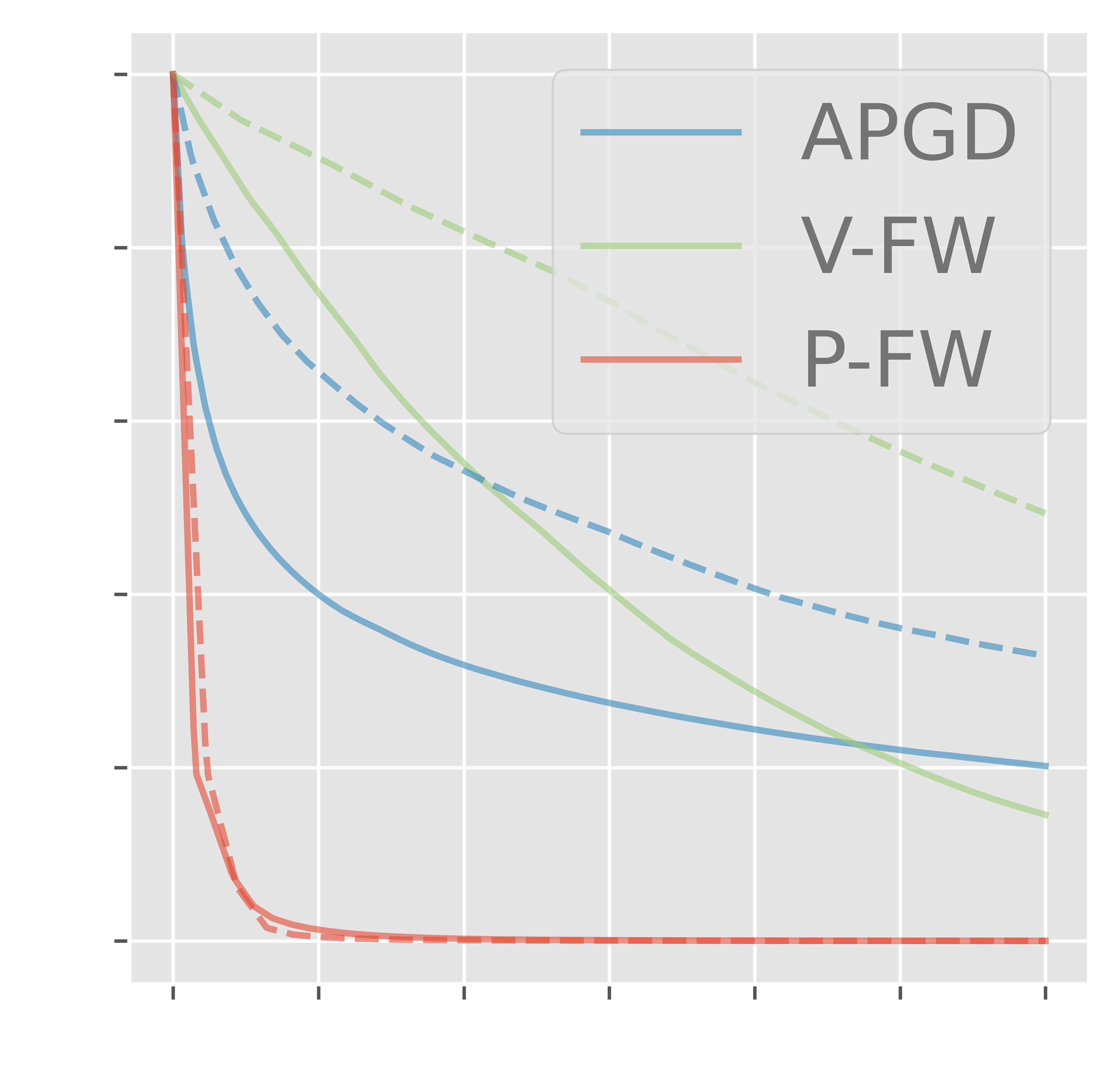}
  }
  
\subfloat[$K=100, N=101^2$]{
  \includegraphics[trim={0 15 0 0},width=.48\linewidth, clip]{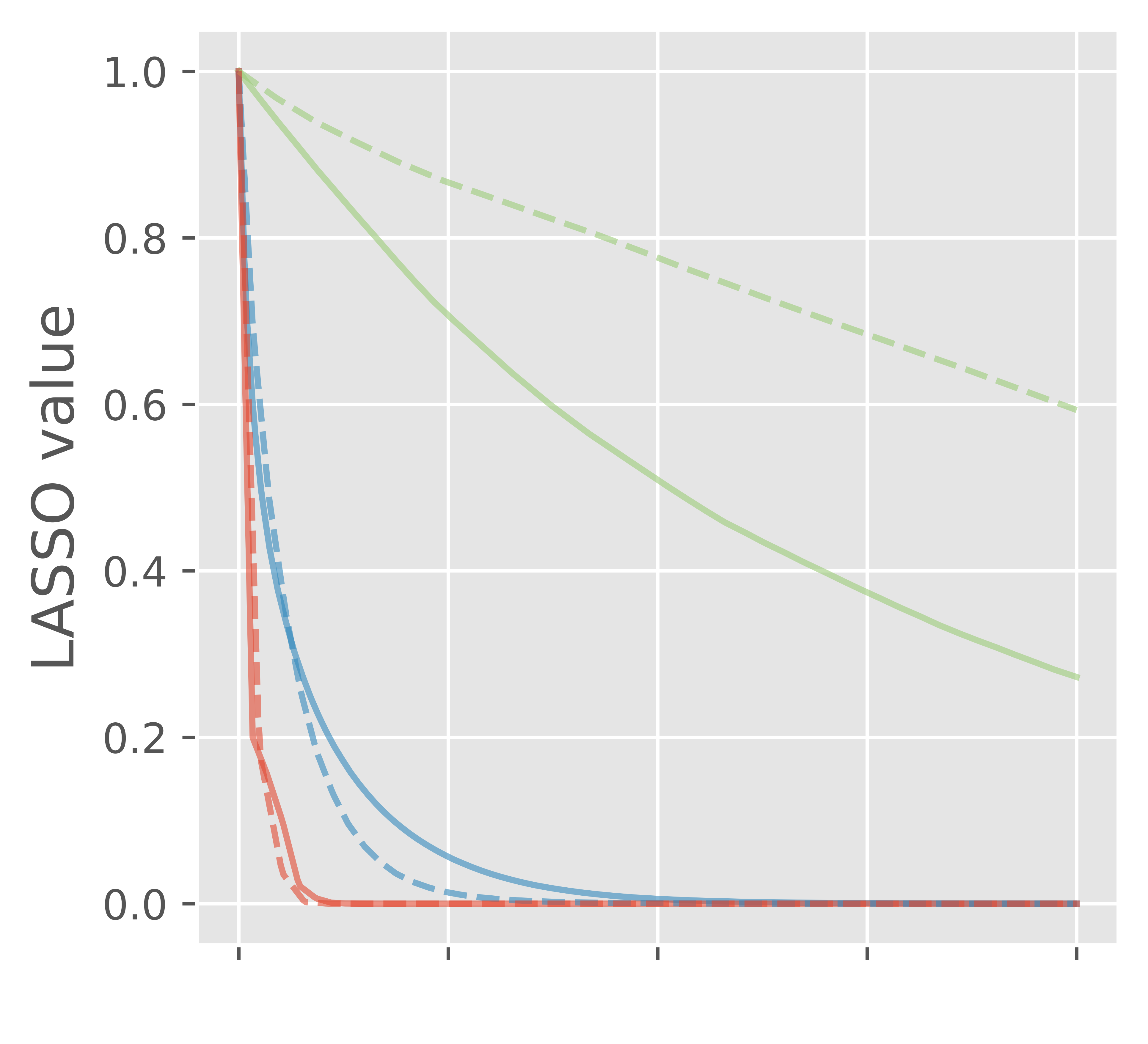}
  }\hfill
\subfloat[$K=100, N=201^2$]{
  \includegraphics[trim={0 15 0 0},width=.45\linewidth, clip]{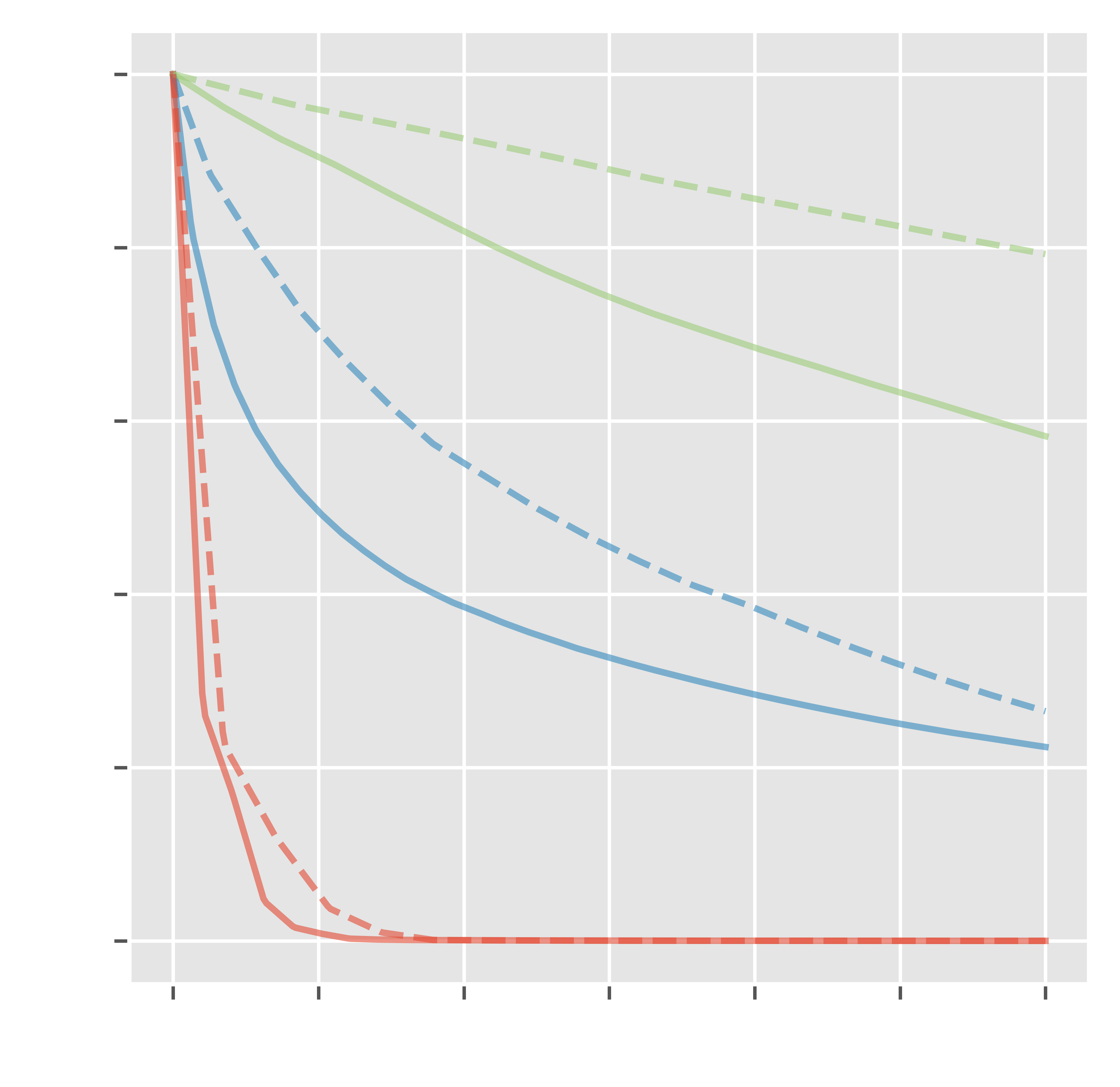}
  }
  
\subfloat[$K=200, N=101^2$ ]{
  \includegraphics[width=.48\linewidth]{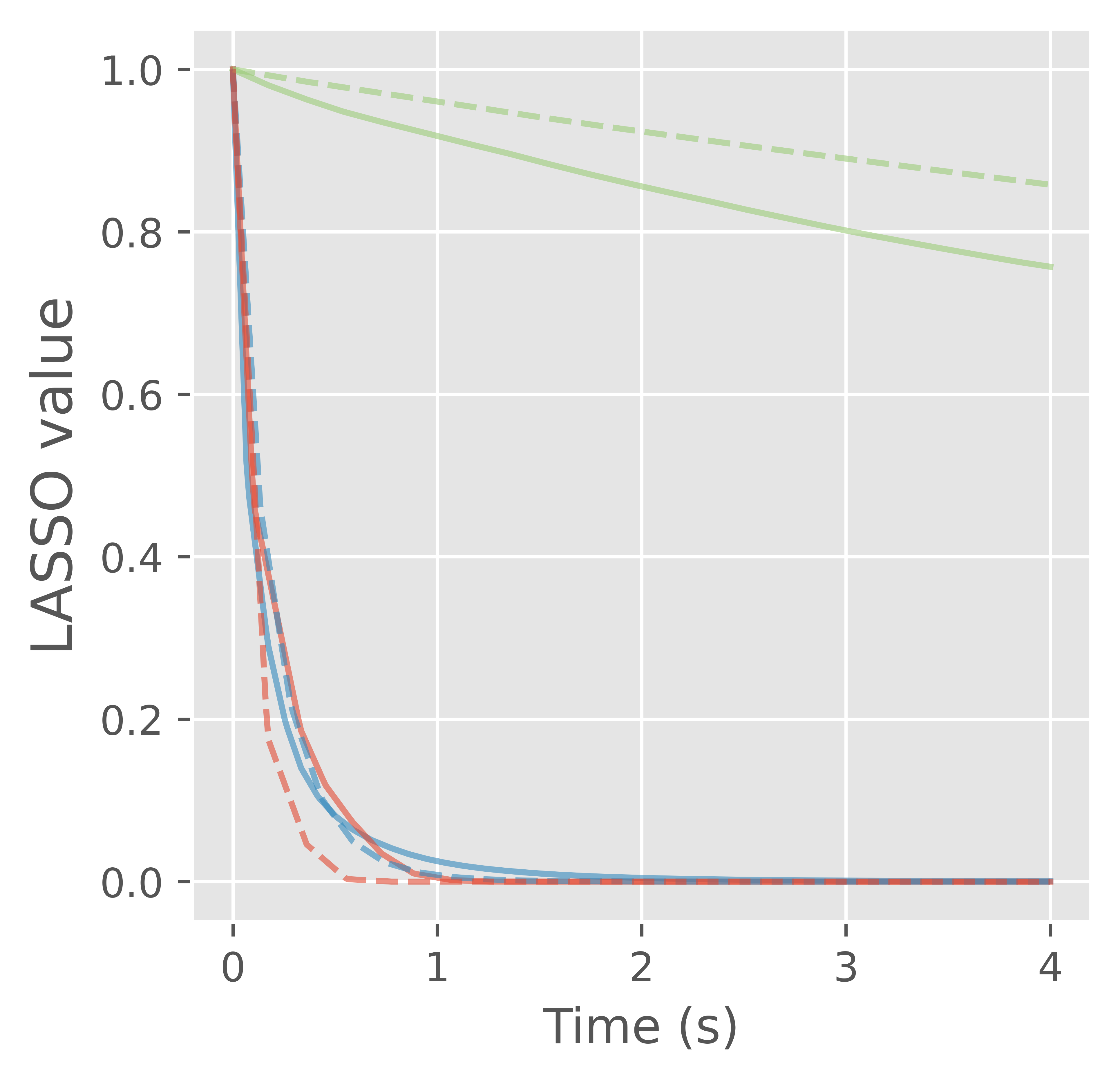}
  }\hfill
\subfloat[$K=200, N=201^2$]{
  \includegraphics[width=.45\linewidth]{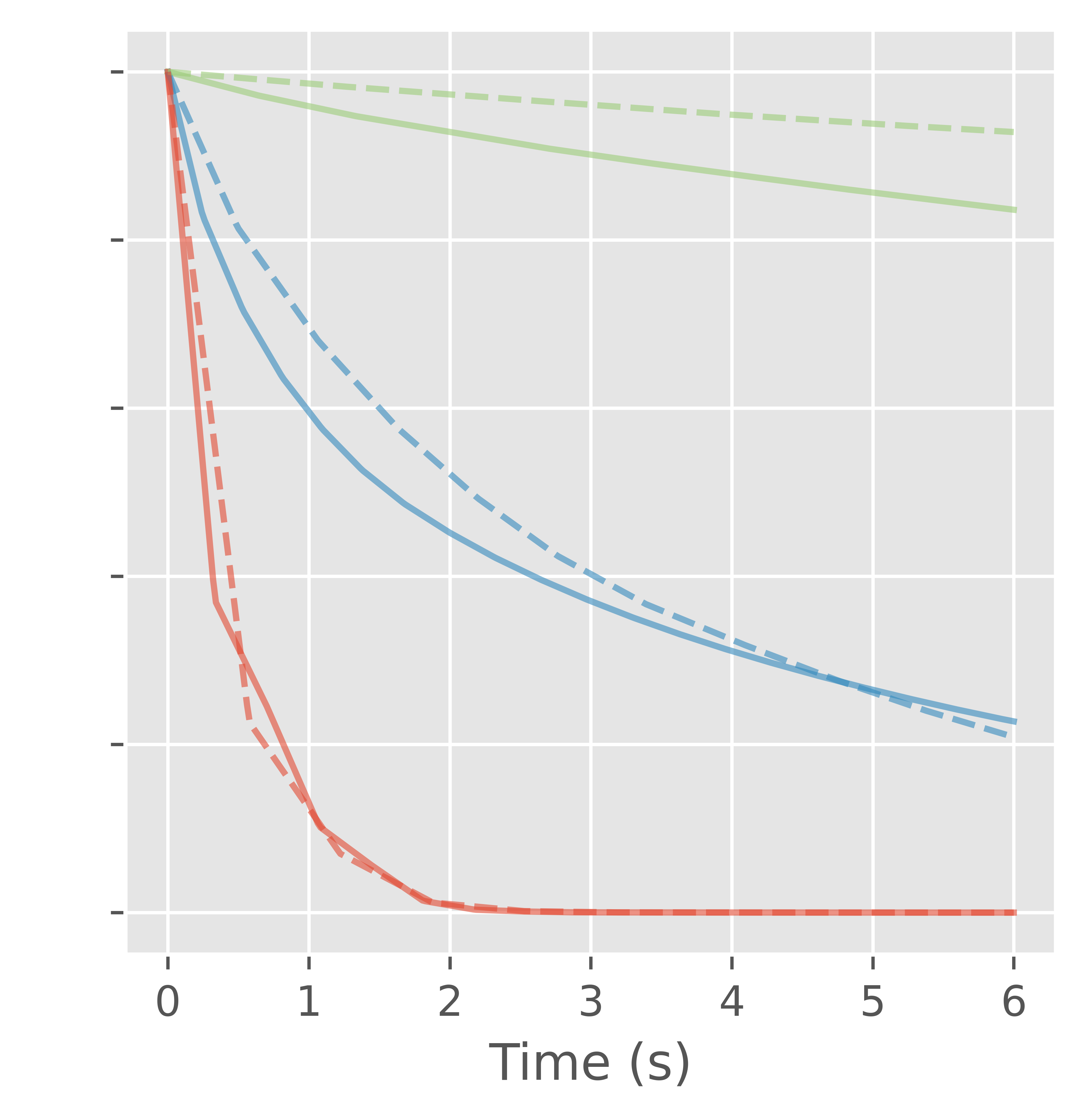}
  }
\caption{Valeur normalisée du problème LASSO au cours du temps de résolution, pour les 3 algorithmes étudiés (APGD, V-FW et P-FW) et différentes tailles d'images $N$ et valeurs de parcimonie sous-jacentes $K$. Les traits pleins et en pointillés correspondent respectivement à $\alpha=8$ et $\alpha=16$. Temps de calculs obtenus avec une machine ThinkPad T14, Intel Core i7 (4C/8T) @ 1.8GHz et 32GB RAM.} 
\label{fig:plots}
\end{center}
\end{figure}

Le contexte expérimental est le suivant. Une fois la matrice de mesure $\mathbf{G}$ calculée selon la procédure détaillée en Section~3, les données sont obtenues selon l'équation $\mathbf{y} = \mathbf{Gx}_0 + \mathbf{w} \in\mathbb{C}^L$, où $\mathbf{x}_0 \in \mathbb{R}^N = \mathbb{R}^{n\times n}$ est une image aléatoire ne comportant que $K$ pixels allumés (indice de parcimonie $K$) et $\mathbf{w}$ est la réalisation d'un vecteur aléatoire gaussien i.i.d. à valeurs complexes (PSNR de $20$dB). Le nombre de mesures est donné par $L = \alpha K$ avec $\alpha=8$ ou $\alpha=16$.  Les résultats numériques sont présentés en Figure~\ref{fig:plots}.

Sur tous les scénarios considérés, V-FW présente toujours les moins bonnes performances, et converge rarement dans les temps impartis. En revanche, la version polyatomique présente d'excellents résultats, et est systématiquement plus rapide que APGD (à l'exception du contexte e), dans lequel les performances sont comparables). Les situations (b) et (d) sont les plus remarquables, P-FW converge en moins de $1$s, alors que APGD n'a atteint que $80$\% de sa marge de progression au bout de $6$s.

Plus en détail, nous constatons que les performances de APGD dépendent peu de la parcimonie $K$ de la source, mais plutôt de la dimension $N$ du problème. Cet algorithme est beaucoup plus long à converger quand l'image d'entrée est de plus grande dimension ($N=201\times 201$) et quand le nombre de mesures augmente ($\alpha=16$ ralentit la convergence en (b), (d) et (f)). À l'inverse, la vitesse de convergence de P-FW semble surtout dépendre de la parcimonie sous-jacente de la solution, l'algorithme étant plus efficace pour des petites valeurs de $K$. Cela peut s'expliquer par son comportement glouton, lui permettant d'identifier efficacement les degrés de liberté actifs de la solution.

\section{Conclusion et perspectives}

Dans ce travail de recherche, nous avons confronté l'algorithme Frank-Wolfe Polyatomique au problème LASSO défini dans un contexte de mesures fréquentielles de Fourier. Tout comme les expériences menées dans \cite{Jarret_Fageot_Simeoni_2022} pour de l'acquisition comprimée, les résultats obtenus démontrent la supériorité de l'algorithme par rapport aux méthodes proximales comme APGD dans les grandes dimensions étudiées. 

Ces résultats   encourageants  représentent une étape intermédiaire dans le but d'appliquer l'algorithme Frank-Wolfe Polyatomique aux données réelles de radioastronomie. Dans cette idée, nous souhaitons développer des opérateurs de mesures de Fourier plus rapides à l'aide de techniques de \textit{Fast Fourier Trasnform (FFT)}. L'algorithme doit également être adapté en conséquence pour manipuler ce type d'opérateurs.



\bibliographystyle{ieeetr}
{\footnotesize
\bibliography{refs}}

\end{document}